\documentclass[10pt,twocolumn]{autart}    

\usepackage{amsmath}
\usepackage{amsfonts}
\usepackage{todonotes}
\usepackage{graphicx}          

\usepackage{epstopdf}
\usepackage[round]{natbib}
\usepackage{pgf,tikz,pgfplots}
\pgfplotsset{compat=1.12}
\usepackage{mathrsfs}
\usepackage{xcolor}
\usetikzlibrary{arrows}
\usetikzlibrary[patterns]

\newtheorem{corollary}{Corollary}
\newtheorem{lemma}{Lemma}
\newtheorem{assumption}{Assumption}
\newtheorem{proposition}{Proposition}
\theoremstyle{definition}
\newtheorem{definition}{Definition}
\newtheorem{problem}{Problem}
\newtheorem{remark}{Remark}

\begin{document}

\begin{frontmatter}

\title{On the confidentiality of controller states\\ under sensor attacks\thanksref{footnoteinfo}} 

\thanks[footnoteinfo]{This work is supported in part by the Swedish Research Council
(grant 2016-00861), the Swedish Energy Agency (project LarGo!), and
the Swedish Civil Contingencies Agency (project CERCES).}

\author[Stockholm]{David Umsonst}\ead{umsonst@kth.se},    
\author[Stockholm]{Henrik Sandberg}\ead{hsan@kth.se},     

\address[Stockholm]{Division of Decision and Control Systems, School of Electrical Engineering and Computer Science, KTH Royal Institute of Technology, Stockholm, Sweden}


\begin{abstract}                          
With the emergence of cyber-attacks on control systems it has become clear that improving the security of control systems is an important task in today's society. We investigate how an attacker that has access to the measurements transmitted from the plant to the controller can perfectly estimate the internal state of the controller. This attack on sensitive information of the control loop is, on the one hand, a violation of the privacy, and, on the other hand, a violation of the security of the closed-loop system if the obtained estimate is used in a larger attack scheme. 
Current literature on sensor attacks often assumes that the attacker has already access to the controller's state.
However, this is not always possible. We derive conditions for when the attacker is able to perfectly estimate the controller's state. These conditions show that if the controller has unstable poles a perfect estimate of the controller state is not possible.
Moreover, we propose a defence mechanism to render the attack infeasible. This defence is based on adding uncertainty to the controller dynamics. We also discuss why an unstable controller is only a good defence for certain plants.
Finally, simulations with a three-tank system verify our results.
\end{abstract}

\end{frontmatter}

\section{Introduction}
The smart grid and intelligent transportation systems are two prime examples of cyber-physical systems, where physical processes are controlled over communication networks and with digital computers. 
The interconnection of the physical and cyber domain promises great advantages in the performance and capabilities of cyber-physical systems.
However, with the introduction of communication networks and computational devices, the controlled processes become vulnerable to cyber-attacks. 
Documented cyber-attacks such as the Stuxnet attack on an Iranian uranium enrichment facility \citep{StuxnetAttack}, the cyber attack on a German steel mill \citep{GermanSteelAttack}, and the BlackEnergy attack on the Ukrainian power grid \citep{UkraineAttack} show that these attacks are not a futuristic concept but already happening.

\citet{SecureControl} define a cyber-physical attack space that is spanned by the attacker's disclosure and disruptive resources as well as its model knowledge. 
Disclosure resources enable the attacker to gather information about the system and, therefore, break its confidentiality. 
These \emph{disclosure attacks} can, for example, be used to increase the attacker's model knowledge.
Disruptive resources, on the other hand, let the attacker launch both deception and denial of service attacks, which affect the integrity and the availability of measurement and actuator signals, respectively.
Several attacks can be mapped into this attack space, for example replay attacks, where well-behaved sensor measurements are replayed, while the actuator signals are changed. 

\begin{figure}
\centering
\includegraphics[scale=0.4]{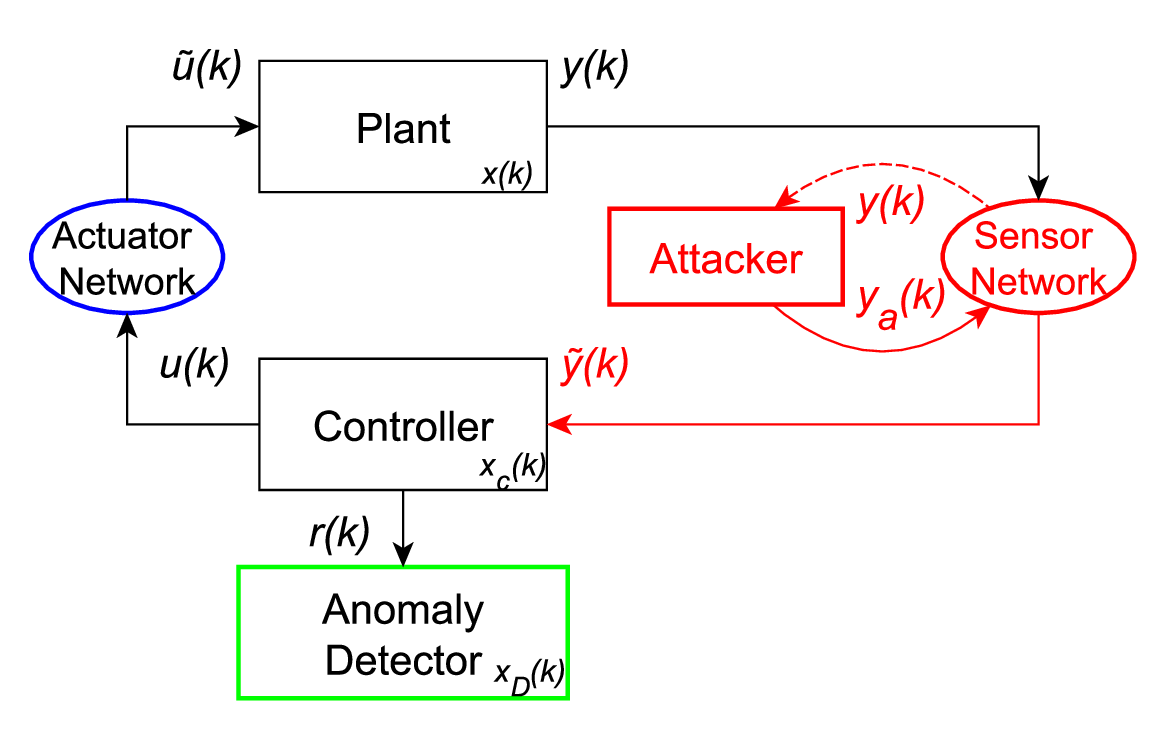}
\caption{Block diagram of the closed-loop system equipped with an anomaly detector under a sensor attack}
\label{fig:BlockDiagSensorAttack}
\end{figure}

Although many attack strategies have been investigated, the analysis of sensor attacks has gained popularity in the last decade. 
A goal of the sensor attacks is to remain undetected by the anomaly detector of the operator, while changing the measurements.
Figure~\ref{fig:BlockDiagSensorAttack} shows the block diagram of the cyber-physical system under a sensor attack.
Here, the dashed line going to the attacker corresponds to the disclosure resources of the attacker.
The disclosure resources can be used to gather more information about the closed-loop system, for example, about the internal states of the plant $x(k)$, the controller $x_c(k)$, or the anomaly detector $x_D(k)$.
The disruptive resources are denoted by $y_a(k)$ and are used to change the values of the measurements from $y(k)$ to $\tilde{y}(k)=y(k)+y_a(k)$. 
\citet{FalseDataInjectionMo} look into integrity attacks on sensors and define a notion of perfectly attackable systems, while \cite{CardenasRiskAssessment} analyse two different detectors and three sensor attack strategies.
Another approach is to maximize the error covariance matrix of a state estimator with a sensor attack as it is done in \cite{LinearSensorAttack}.
In \citet{RuthsMultivariate}, a sensor attack strategy is proposed which replaces the residual signal $r(k)$, which is the input to the anomaly detector (Fig.~\ref{fig:BlockDiagSensorAttack}), with a signal designed by the attacker. 
This attack strategy is then used to look at the impact under the two anomaly detectors investigated in \cite{CardenasRiskAssessment}.
In \citet{UmsonstACC19}, it was shown how an attacker using this attack strategy is able to break the confidentiality of the internal anomaly detector state $x_D(k)$.

What connects all these papers on sensor attacks is that the attacker needs to have \emph{exact} knowledge about the internal state $x_c(k)$ of the controller, when the attack starts.
\citet{FalseDataInjectionMo} assume that the initial system state is zero in order to determine the undetectable attack, while the other papers assume that the controller state is known to the attacker.
Therefore, this paper investigates a missing piece that is often taken for granted in sensor attacks, namely the broken confidentiality of the controller's internal state. 
More precisely, we examine if an attacker with full model knowledge listening to the sensor measurements is able to break the confidentiality of the controller.
This confidentiality attack can have two purposes. 
One purpose might be that the attacker is curious and wants to follow the activity of the control centre. 
The other purpose might be that it is one step in a more complex attack scheme. 
This step represents gathering information of the plant and its controller, which is then used in later steps to attack the system. 
We can interpret this as a first step the attacker needs to perform to execute the attacks proposed by the papers mentioned above.

\subsection{Contributions}
The contribution of our work is three-fold.
Firstly, we provide a rigorous analysis of whether or not an attacker with full-model knowledge and access to all sensors is able to \emph{perfectly} estimate the internal state of the output-feedback controller. 
Although it may seem obvious that such a powerful attacker is able to estimate the state perfectly, we show that if the operator uses an unstable controller the attacker is not able to do so. 
We further classify all gains for a linear time-invariant observer the attacker could use to achieve a perfect estimate in case of a stable output-feedback controller.
The second contribution provides a defence mechanism against this disclosure attack. 
This mechanism proposes to add some uncertainty to the controller's input, which can be interpreted as a watermarking scheme. 
Furthermore, we discuss when an unstable controller is an appropriate defence mechanism.
The third contribution is the verification of our theoretical results with simulations of a three-tank system under attack.
	
\subsection{Related work}

Most of the research on the security of cyber-physical systems has focused on the integrity and availability of data, according to \citet{AttackSurvey}. For example, all the previously mentioned papers on sensor attacks except \citet{UmsonstACC19} consider integrity attacks. 

Other work on the confidentiality of control systems can be found in, for example, \citet{ConfidentialityInVehicleNetworks}, \citet{MoConfidentialityOfControllerStructure}, and \citet{ConfidentialityOfWideAreaPowerSystems}.
What distinguishes this paper from these results is that we focus on a general linear system structure, while \citet{ConfidentialityInVehicleNetworks} investigate the confidentiality of a special structured linear system.
Further, we focus on the confidentiality of the controller's internal state.
However, \citet{ConfidentialityInVehicleNetworks} consider the confidentiality of the whole system state, while \citet{ConfidentialityOfWideAreaPowerSystems} look into the confidentiality of controller gains and in \citet{MoConfidentialityOfControllerStructure} the attacker wants to identify the controller structure. 
In \citet{MoConfidentialityOfControllerStructure}, it is shown that an appropriate controller design can lead to confidentiality. We also show that a certain type of controller, in our case an unstable controller, leads to confidentiality regarding sensitive information of the controller. It is interesting that an appropriate controller design can preserve the controller's confidentiality.

Another recent research direction is to use homomorphic encryption to ensure the security and privacy of control systems \citep{HomomorphicEncryptionCDC15,HomomorphicEncryptionFarokhi}. 
Based on encrypted sensor measurements the controller determines an encrypted control signal, which is decrypted at the actuator, i.e., the feedback loop operates on encrypted signals.
The use of encrypted signals guarantees that, even if the attacker estimates the controller state, the estimate is not useful to the attacker, due to the encryption.
In our approach, we use an artificial uncertainty instead of encryption techniques to preserve the confidentiality of the controller.

Defending the cyber-physical system against attacks by introducing an artificial uncertainty is also done in the work using watermarking. 
Watermarking of the actuator signal has been considered as a defence mechanism, for example, against replay attacks \citep{MoWatermarking} or sensors attacks in networked control systems \citep{WatermarkingNetworked}. 
However, in this paper, the uncertainty is added to the input of the controller, while watermarking techniques usually add it to the output of the controller, i.e., the actuator signal.
\subsection{Notation}
Let $x$ be a column vector in $\mathbb{R}^n$ and $A$ a matrix in $\mathbb{R}^{n\times m}$.
The spectral radius of a square matrix $A$ is $\rho(A)$. 
Further, we say $A$ is (Schur) stable, if ${\rho(A)<1}$.
The trace of $A$ is denoted as $\mathrm{tr}(A)$.
By ${B>0}$ ($B\geq 0$), we mean a matrix is symmetric positive definite (semi-definite).
The identity matrix of dimension $n$ is denoted as $I_n$, while $0$ denotes either a scalar, a vector, or a matrix with all elements equal to zero. The dimension of $0$ is clear from the context. A Gaussian random variable $x$ with mean $\mu$ and covariance matrix $\Sigma$ is denoted as $x\sim\mathcal{N}(\mu,\Sigma)$.

\section{Problem formulation}
In this section, we present the models of the plant and controller. Further, we describe the assumptions on and the goals of the attacker, which set the stage for the formulation of the problem.
\subsection{Plant and controller model}
The plant is modelled as a linear discrete-time system,
\begin{equation}
\label{eq:PlantDynamics}
	\begin{aligned}
		x(k+1)&=Ax(k)+Bu(k)+w(k),\\
		y(k)&=Cx(k)+v(k),
	\end{aligned}
\end{equation}
where $x(k)$ is the state of the plant in $\mathbb{R}^{n_x}$, $u(k)$ is the plant input in $\mathbb{R}^{n_u}$, and $y(k)$ is the measured output in $\mathbb{R}^{n_y}$. 
Further, $A\in\mathbb{R}^{n_x \times n_x}$ is the system matrix, $B\in\mathbb{R}^{n_x\times n_u}$ is the input matrix, and $C\in\mathbb{R}^{n_y \times n_x}$ is the output matrix. 
Here, $w(k)\sim\mathcal{N}(0,\Sigma_w)$ is the process noise and $v(k)\sim\mathcal{N}(0,\Sigma_v)$ is the measurement noise, where $\Sigma_w\geq 0$ and $\Sigma_v>0$ are the covariance matrices of the respective noise terms and have appropriate dimensions. 
The noise processes $w(k)$ and $v(k)$ are each independent and mutually uncorrelated.
The operator uses an output-feedback controller of the form
\begin{equation}
\label{eq:ControllerDynamics}
	\begin{aligned}
	x_c(k+1)&=A_cx_c(k)+B_c y(k),\\
	u(k)&=C_cx_c(k)+D_cy(k),
	\end{aligned}
\end{equation}
where $x_c(k)$ is the controller's state in $\mathbb{R}^{n_c}$, $A_c\in\mathbb{R}^{n_c \times n_c}$ is the system matrix of the controller, $B_c\in\mathbb{R}^{n_c \times n_y}$ is the input matrix of the controller, $C_c\in\mathbb{R}^{n_u \times n_c}$ is the output matrix of the controller, and $D_c\in\mathbb{R}^{n_u \times n_y}$ is the feedthrough matrix from the measurements to the actuator signal.
This structure can represent many commonly used controllers. 
For example, with $A_c=A-BK-LC$, $B_c=L$, $C_c=-K$, and $D_c=0$, we obtain an observer-based controller, where $x_c(k)$ is an estimate of $x(k)$, and $K$ and $L$ represent the feedback and observer gain, respectively. The observer-based controller is, for example, used in \cite{RuthsMultivariate}.

The closed-loop system dynamics can be written as
\begin{multline*}
	\begin{bmatrix}
	x(k+1)\\
	x_c(k+1)
	\end{bmatrix}=\begin{bmatrix}
	A+BD_cC &BC_c\\
	B_cC &A_c
	\end{bmatrix}
	\begin{bmatrix}
	x(k)\\ x_c(k)
	\end{bmatrix}\\
	+\begin{bmatrix}
	w(k)+BD_cv(k)\\ B_cv(k)
	\end{bmatrix}.
\end{multline*}
Introducing 
\begin{align*}
z(k)=\begin{bmatrix}
x(k)\\ x_c(k)
\end{bmatrix}\ \mathrm{and}\ \eta^\prime(k)=\begin{bmatrix}
w(k)+BD_cv(k)\\ B_cv(k)
\end{bmatrix}
\end{align*}
we write the closed-loop system as
\begin{equation}
\label{eq:ClosedLoopDynamicsWithCorrelatedNoise}
\begin{aligned}	
		z(k+1)&=A_z^\prime z(k)+\eta^\prime(k)\\
		y(k)&=C_zz(k)+v(k)=\begin{bmatrix}
		C &0
		\end{bmatrix}z(k)+v(k),
\end{aligned}
\end{equation}
where $\eta^\prime(k)\sim\mathcal{N}(0,Q^\prime)$ is the zero mean process noise of the closed-loop system with covariance matrix ${Q^\prime\in\mathbb{R}^{(n_x+n_c)\times (n_x+n_c)}}$ and $v(k)$ is the measurement noise. 

\begin{assumption}
\label{assum:ControlObserverAssumptions}
	The system is such that
	\begin{enumerate}
		\item  $(A,B)$ is stabilizable,
		\item  $(C,A)$ is detectable,
		\item  $(A,\Sigma_w^{\frac{1}{2}})$ has no uncontrollable modes on the unit circle, and
		\item  the controller $(A_c,B_c,C_c,D_c)$ is minimal.
	\end{enumerate}
\end{assumption}

The stability of $A_z^\prime$ depends on the controller matrices $A_c,B_c,C_c$, and $D_c$. Therefore, we need the first two points of Assumption~\ref{assum:ControlObserverAssumptions} such that the operator is able to observe and control all unstable modes in the system. 
The third point is needed later for the existence of the solution of a Riccati equation.
To avoid unnecessary dynamics, the implementation of the controller should be its minimal realization.
\begin{assumption}
\label{assum:StableClosedLoop}
The operator has designed $A_c,B_c,C_c$, and $D_c$, such that the closed-loop system is stable, i.e., $\rho(A_z^\prime)<1$.
\end{assumption}
Assuming a stable closed-loop system is in line with normal operator requirements.

\begin{assumption}
\label{assum:SteadyStateClosedLoop}
The closed-loop system has reached steady state before $k=0$ and $z(0)\sim\mathcal{N}(0,\Sigma_0)$, where $\Sigma_0\geq 0$ is the solution to
\begin{align*}
\Sigma_{0}=A_z^\prime\Sigma_{0}(A_z^\prime)^T+Q^\prime.
\end{align*}
\end{assumption}
This assumption is not restrictive, since industrial plants usually run for long periods of time, and we know that the covariance of $z(k)$ will reach its unique steady state, since $\rho(A_z^\prime)<1$ by Assumption~\ref{assum:StableClosedLoop}.

Note that the closed-loop process noise variable $\eta^\prime(k)$ is correlated with the measurement noise $v(k)$,
\begin{align*}
\mathbb{E}&\bigg\lbrace\begin{bmatrix}
\eta^\prime(k) \\ v(k)
\end{bmatrix}\begin{bmatrix}
\eta^\prime(k)^T & v(k)^T
\end{bmatrix}\bigg\rbrace\\
&=\left[
\begin{array}{cc|c}
\Sigma_w+BD_c\Sigma_vD_c^TB^T & BD_c\Sigma_vB_c^T & BD_c\Sigma_v\\ B_c\Sigma_vD_c^TB^T & B_c\Sigma_vB_c^T & B_c\Sigma_v\\ \hline \Sigma_vB^TD_c^T & \Sigma_v^TB_c^T & \Sigma_v
\end{array}
\right]\\
&=\left[\begin{array}{c|c}
Q^\prime &S\\ \hline S^T &R
\end{array}\right],
\end{align*}
where $S\in\mathbb{R}^{(n_x+n_c)\times n_y}$, and ${R\in\mathbb{R}^{n_y\times n_y}}$.

Since the $\eta^\prime(k)$ and $v(k)$ are correlated, we will apply a transformation proposed in \citet{ConvergenceOfRiccatiDifferenceEquation} to obtain a system representation with uncorrelated noises.
\begin{align*}
z(k+1)&=A_z^\prime z(k)+\eta^\prime(k)-SR^{-1}(y(k)-y(k))\\
&=A_zz(k)+\eta(k)+SR^{-1}y(k),
\end{align*}
where $A_z=A_z^\prime-SR^{-1}C_z$,
\begin{align*}
\eta(k)=\eta^\prime(k)-SR^{-1}v(k)=\begin{bmatrix}
w(k)\\ 0
\end{bmatrix},
\end{align*}
\begin{align*}
\mathbb{E}&\bigg\lbrace\begin{bmatrix}
\eta(k) \\ v(k)
\end{bmatrix}\begin{bmatrix}
\eta(k)^T & v(k)^T
\end{bmatrix}\bigg\rbrace=\left[\begin{array}{c|c}
Q &0\\ \hline 0 &R
\end{array}\right],
\end{align*}
and
\begin{align*}
Q=Q^\prime-SR^{-1}S^T=\begin{bmatrix}
\Sigma_w & 0\\0 & 0
\end{bmatrix}.
\end{align*}
The zero elements in $Q$ show us that there is no process noise acting on the controller in the transformed system.

Therefore, the closed-loop dynamics we consider from now on are
\begin{equation}
\begin{aligned}
z(k+1)&=A_z z(k)+\eta(k)+SR^{-1}y(k),\\
y(k)&=C_zz(k)+v(k).
\end{aligned}
\label{eq:closedLoopDynamics}
\end{equation}
Note that even though $\rho(A_z^\prime)<1$, it is not always the case that $\rho(A_z)<1$. 

\subsection{Attack model and goals}
Now that we introduced the plant and controller model, we look into the attack model and the attacker's goal. We begin by introducing the assumptions made about the attacker.
\begin{assumption}
\label{assum:AttackerKnowledge}
	The attacker has gained access to the model $(A,B,C,A_c,B_c,C_c,D_c)$, the noise statistics $(\Sigma_w, \Sigma_v)$, the measurements $y(k)$ for $k\geq 0$ but \emph{not} the control signals $u(k)$ and the initial state of the system $z(0)$. 
\end{assumption}
Since the manipulation of control signals can lead to an immediate physical impact, we assume $u(k)$ is better protected and therefore the attacker does not have access to it. 
Moreover, we set the start of the attack arbitrarily to $k=0$. This can be interpreted as the point in time, from which the attacker has access to the measurements. 
From Assumption~\ref{assum:SteadyStateClosedLoop} we know that the plant and controller have been running for a long time. Therefore, the attacker does not know the state $z(0)$ when it gains access to the sensor measurements.

\begin{assumption}
\label{assum:TimelyEstimate}
The attacker uses measurements up to time step $k$ to estimate the controller's internal state at time step $k+1$.
\end{assumption}
It is possible to use measurements up to time step ${k^*\geq k+1}$ to estimate the controller's state at time step $k+1$. 
However, if the attacker wants to launch a false-data injection attack at time step $k+1$, this estimate needs to be available already.

The \emph{goal} of the attacker is to obtain an estimate $\hat{x}_c(k)$, such that this estimate perfectly tracks the controller state $x_c(k)$ as $k$ grows large. 
This can be either a first step in a larger attack scheme or a way to gain some insight in the controller's internal state.
The goal can be formulated as the following problem.
\begin{problem}
\label{prob:ControllerStateEstimation}
Estimate $x_c(k)$ such that the estimation error is unbiased, i.e., $\mathbb{E}\lbrace x_c(k)-\hat{x}_c(k)\rbrace=0$, and its covariance matrix $\Sigma_c(k)$ approaches zero, i.e.,
\begin{align*}
\lim_{k\rightarrow\infty}\Sigma_c(k)=0
\end{align*}
for a given $\Sigma_c(0)\geq 0$.
\end{problem}
An estimation error covariance matrix $\Sigma_c(k)$ that approaches zero as $k$ grows large means the estimate converges to the true value in mean square (and thus also in probability).

Note that the controller has a minimal realization (see Assumption~\ref{assum:ControlObserverAssumptions}), and having access to both actuator and measurement signals would mean we can always perfectly estimate its state using a standard observer involving only the controller model.

In Section~\ref{sec:AttackStrategy} we characterize for which systems the controller's confidentially can be broken (Problem~\ref{prob:ControllerStateEstimation}), and in Section~\ref{sec:DefenceMechanism} we discuss possible defence mechanisms.

\section{Estimating the controller's state $x_c(k)$}
\label{sec:AttackStrategy}
In this section, we investigate when a solution to Problem~\ref{prob:ControllerStateEstimation} exists. 
It may seem obvious that an attacker according to Assumption~\ref{assum:AttackerKnowledge} is without any doubt able to estimate the controller's state $x_c(k)$ perfectly. 
However, we show in the following that this is not always the case.
First, we present the optimal attack strategy to estimate $x_c(k)$ and then state conditions for the convergence of $\Sigma_c(k)$ to zero. 
Following this, we look into non-optimal strategies to solve Problem~\ref{prob:ControllerStateEstimation}.

\subsection{Optimal attack strategy}
To obtain the optimal attack strategy, we start by investigating the conditional probability of the closed-loop system state $z(k+1)$ given all measurements up to time step $k$.
Due to the presence of the process noise, $\eta(k)$, and measurement noise, $v(k)$, we know that $z(k+1)$ is a random variable. 
Since \eqref{eq:closedLoopDynamics} is a linear system with Gaussian noise, we know that $z(k+1)$ given the measurements up to time step $k$ is also a Gaussian random variable \citep{OptimalFiltering}.
Let $\lbrace y(i)\rbrace_{i=0}^{l}$ be the sequence $\{y(0),\cdots,y(l)\}$, then the conditional probability distribution of $z(k+1)$ given $\{y(i)\}_{i=0}^{k}$ is
\begin{align*}
	z(k+1|\{y(i)\}_{i=0}^{k})\sim\mathcal{N}\big(\hat{z}(k+1),\Sigma_z(k+1)\big),
\end{align*}
where
\begin{align}
	\label{eq:KalmanFilterAttacker}
	\hat{z}(k+1)=A_z\hat{z}(k)+SR^{-1}y(k)+L_z(k)\big(y(k)-C_z\hat{z}(k)\big)
\end{align}
is the conditional mean of $z(k+1)$ with $L_z(k)=\big(A_z\Sigma_z(k)C_z^T\big)\big(C_z\Sigma_z(k)C_z^T+R\big)^{-1}$, $\hat{z}(0)=\mathbb{E}\lbrace z(0)\rbrace=0$, and
\begin{equation}
\label{eq:RiccatiOneStepAhead}
	\begin{aligned}
		&\Sigma_z(k+1)=A_z\Sigma_z(k)A_z^T+Q\\
		&-\big(A_z\Sigma_z(k)C_z^T\big)\big(C_z\Sigma_z(k)C_z^T+R\big)^{-1}\big(A_z\Sigma_z(k)C_z^T\big)^T
	\end{aligned}
\end{equation}
is the conditional covariance matrix. Its initial condition is ${\Sigma_z(0)=\Sigma_{0}}$, which is given in Assumption~\ref{assum:SteadyStateClosedLoop}.

The optimal estimator for $z(k)$ given $\{y(i)\}_{i=0}^{k}$ is the Kalman filter \citep{OptimalFiltering}. It is optimal in the sense that it minimizes the mean square error. 
Therefore, the \emph{optimal} attack strategy to estimate $x_c(k)$ is a time-varying Kalman filter, which uses $\hat{z}(k)$ in \eqref{eq:KalmanFilterAttacker} as the estimate of $z(k)$.
The goal of the attacker is to have an estimate $\hat{z}(k)$ of the closed-loop system's state such that $\begin{bmatrix}
0 &I_{n_c}
\end{bmatrix}\hat{z}(k)\rightarrow x_c(k)$ as $k\rightarrow \infty$.

Instead of directly analysing $\hat{z}(k)$, we introduce the estimation error $e_z(k)=z(k)-\hat{z}(k)$ that has the dynamics 
\begin{align*}
	e_z(k+1)=\big(A_z-L_z(k)C_z\big)e_z(k)+\eta(k)+L_z(k)v(k).
\end{align*}
and covariance matrix
\begin{align*}
	\mathbb{E}\big\lbrace e_z(k+1)e_z(k+1)^T \big\rbrace=\Sigma_z(k+1).
\end{align*}
A Kalman filter is an unbiased estimator, which means that $\mathbb{E}\lbrace z(k)\rbrace=\hat{z}(k)$, or, differently formulated, $\mathbb{E}\lbrace e_z(k)\rbrace=0$. 
Hence, Problem~\ref{prob:ControllerStateEstimation} is solved if, for $\Sigma_z(0)=\Sigma_{0}$, the attacker's Kalman filter fulfils
\begin{align}
\label{eq:DesiredCovarianceMatrix}
	\lim_{k\rightarrow\infty}\Sigma_{z}(k)=\begin{bmatrix}
	P &0\\0 &0
	\end{bmatrix},
\end{align}
where $P\geq 0$. 
Note that $\Sigma_{0}$ can be calculated by the attacker because of its model knowledge by Assumption~\ref{assum:AttackerKnowledge}.

\subsection{Asymptotic convergence to $\Sigma_c(k)=0$}
Let us now investigate when the optimal attack strategy solves Problem~\ref{prob:ControllerStateEstimation}. Here, we present necessary and sufficient conditions for the covariance matrix $\Sigma_c(k)$ to converge to zero. 
Recall this is equivalent to saying that \eqref{eq:DesiredCovarianceMatrix} is fulfilled.

Before we present our convergence results, note that a steady state solution to \eqref{eq:RiccatiOneStepAhead} satisfies the algebraic Riccati equation (ARE)
\begin{multline}
\label{eq:RiccatiOneStepAheadARE}
		\Sigma_\infty=A_z\Sigma_\infty A_z^T+Q\\
		-\big(A_z\Sigma_\infty C_z^T\big)\big(C_z\Sigma_\infty C_z^T+R\big)^{-1}\big(A_z\Sigma_\infty C_z^T\big)^T,
\end{multline}
where $L_\infty=\big(A_z\Sigma_\infty C_z^T\big)\big(C_z\Sigma_\infty C_z^T+R\big)^{-1}$ is the steady state Kalman gain.
\begin{definition}[Definition 3.1 \citep{ConvergenceOfRiccatiDifferenceEquation}]
A real symmetric nonnegative definite solution $\Sigma_\infty$ to \eqref{eq:RiccatiOneStepAheadARE} is called a \emph{strong} solution if $\rho(A_z-L_\infty C_z)\leq 1$. The strong solution is called a \emph{stabilizing} solution if $\rho(A_z-L_\infty C_z)<1$.
\end{definition}
The following lemma from \citet{ConvergenceOfRiccatiDifferenceEquationWithSingularMatrices} will be useful in the following discussion.
\begin{lemma}[Theorem 3.2 \citep{ConvergenceOfRiccatiDifferenceEquationWithSingularMatrices}]
\label{lem:ConditionsForARESolutions}
Let $G^TG=Q$,
\begin{enumerate}
\item the strong solution of the ARE exists and is unique if and only if $(C_z,A_z)$ is detectable;
\item the strong solution is the only nonnegative definite solution of the ARE if and only if $(C_z,A_z)$ is detectable and $(A_z,G)$ has no uncontrollable modes outside the unit circle;
\item the strong solution coincides with the stabilizing solution if and only if $(C_z,A_z)$ is detectable and $(A_z,G)$ has no uncontrollable modes on the unit circle;
\item the stabilizing solution is positive definite if and only if $(C_z,A_z)$ is detectable and $(A_z,G)$ has no uncontrollable modes inside, or on the unit circle.
\end{enumerate}
\end{lemma}
Let us begin by showing that a solution to \eqref{eq:RiccatiOneStepAheadARE} of the form in \eqref{eq:DesiredCovarianceMatrix} exists.
\begin{proposition}
\label{prop:SteadyStateSolutionOfDesiredForm}
A solution of the algebraic Riccati equation~\eqref{eq:RiccatiOneStepAheadARE} is given by
\begin{align*}
	\Sigma_{\infty}=\begin{bmatrix}
	P &0\\0 &0
	\end{bmatrix},
\end{align*}
where $P\geq 0$ is the unique solution of the ARE
\begin{align*}
P=APA^T+\Sigma_w-APC^T(CPC^T+\Sigma_v)^{-1}CPA^T.
\end{align*}
\end{proposition}
\begin{pf}
Let us first determine
\begin{align*}
A_z=A_z^\prime-SR^{-1}C_z=\begin{bmatrix}
A & BC_c\\ 0 & A_c
\end{bmatrix}.
\end{align*}
After algebraic computations we obtain
\begin{multline*}
\!\!\!\!\!\!\!\!\!\! A_z\Sigma_{\infty}A_z^T+Q=
	\begin{bmatrix}
	APA^T+\Sigma_w & 0\\
	0 & 0
	\end{bmatrix},
A_z\Sigma_{\infty}C_z^T=\begin{bmatrix}
	APC^T\\
	0
	\end{bmatrix},
\end{multline*}
and $C_z\Sigma_{\infty}C_z^T+R=CPC^T+\Sigma_v$
such that
\begin{align*}
	&\big(A_z\Sigma_{\infty}C_z^T\big)\big(C_z\Sigma_{\infty}C_z^T+R\big)^{-1}\big(A_z\Sigma_{\infty}C_z^T\big)^T=\\
	&\begin{bmatrix}
	APC^T(CPC^T+\Sigma_v)^{-1}CPA^T & 0\\
	0 &0
	\end{bmatrix}.
\end{align*}
This leads to
\begin{align*}
\!\!\!\! \Sigma_{\infty}=
	\begin{bmatrix}
	APA^T+\Sigma_w-APC^T(CPC^T+\Sigma_v)^{-1}CPA^T &0\\ 0 & 0
	\end{bmatrix}.
\end{align*}
For $\Sigma_{\infty}$ to be a  solution of \eqref{eq:RiccatiOneStepAheadARE} we require
\begin{align}
\label{eq:RiccatiOperatorOneStep}
P=APA^T+\Sigma_w-APC^T(CPC^T+\Sigma_v)^{-1}CPA^T.
\end{align}
Note that \eqref{eq:RiccatiOperatorOneStep} by itself is an algebraic Riccati equation. 
It is actually the algebraic Riccati equation an operator would obtain when it is designing a time-invariant Kalman filter.
Due to detectability of $(C,A)$ (Assumption~\ref{assum:ControlObserverAssumptions}), there exists a unique strong solution $P\geq 0$ for \eqref{eq:RiccatiOperatorOneStep} (Lemma~\ref{lem:ConditionsForARESolutions}).
Hence, $\Sigma_{\infty}$ is a solution of \eqref{eq:RiccatiOneStepAheadARE}.
\end{pf}
Now that we proved that $\Sigma_{\infty}$ is indeed a solution to the algebraic Riccati equation, we need to show under which conditions $\Sigma_z(k)$ converges to $\Sigma_{\infty}$ for the initial condition $\Sigma_{0}$.

\begin{lemma}
\label{lem:StrongSolutionIffControllerModesInsideAndOnUnitCircle}
The unique strong solution of the ARE~\eqref{eq:RiccatiOneStepAheadARE} is $\Sigma_{\infty}$ if and only if $\rho(A_c)\leq 1$.
\end{lemma}
\begin{pf}
Due to the first statement in Lemma~\ref{lem:ConditionsForARESolutions}, the strong solution is unique and exists if and only if $(C_z,A_z)$ is detectable.
From the stability of $A_z^\prime=A_z+SR^{-1}C_z$, it follows that $(C_z,A_z)$ is detectable.
Hence, the strong solution will be unique.
Further, if $\rho(A_z-L_{\infty}C_z)\leq 1$ for 
\begin{align*}
L_{\infty}&=\big(A_z\Sigma_{\infty}C_z^T\big)\big(C_z\Sigma_{\infty}C_z^T+R\big)^{-1}\\
&=\begin{bmatrix}
APC^T(CPC^T+\Sigma_v)^{-1}\\
0
\end{bmatrix}=\begin{bmatrix}
\bar{L}\\0
\end{bmatrix},
\end{align*}
then $\Sigma_{\infty}$ is a strong solution.
Let us now look at the eigenvalues of $A_z-L_{\infty}C_z$, which are determined by the eigenvalues of $A-\bar{L}C$ and $A_c$, because
\begin{align*}
A_z-L_{\infty}C_z=\begin{bmatrix}
A-\bar{L}C &BC_c\\
0 & A_c
\end{bmatrix}.
\end{align*}
Due to the detectability of $(C,A)$ (Assumption~\ref{assum:ControlObserverAssumptions}), the first statement of Lemma~\ref{lem:ConditionsForARESolutions} shows us that $P$ is a strong solution of \eqref{eq:RiccatiOperatorOneStep}, such that $\rho(A-\bar{L}C)\leq 1$.
Therefore, $\rho(A_z-L_{\infty}C_z)\leq 1$, i.e., $\Sigma_{\infty}$ is the unique strong solution, if and only if $\rho(A_c)\leq 1$.
\end{pf}
\begin{thm}
\label{thm:ConvergenceToSrongSolution}
The covariance matrix $\Sigma_z(k)$ converges to the attacker's desired covariance matrix $\Sigma_{\infty}$ for the initial condition $\Sigma_{0}$ if and only if $\rho(A_c)\leq 1$.
\end{thm}
\begin{pf}
By Lemma~\ref{lem:StrongSolutionIffControllerModesInsideAndOnUnitCircle}, $\Sigma_{\infty}$ is the unique strong solution of \eqref{eq:RiccatiOneStepAheadARE} if and only if $\rho(A_c)\leq 1$.
Theorem 4.2 in \citet{ConvergenceOfRiccatiDifferenceEquationWithSingularMatrices} states that subject to $\Sigma_{0}-\Sigma_{\infty}\geq 0$ the covariance matrix $\Sigma_z(k)$ will converge to the strong solution $\Sigma_{\infty}$ if and only if $(C_z,A_z)$ is detectable.
That $(C_z,A_z)$ is detectable is shown in the proof of Lemma~\ref{lem:StrongSolutionIffControllerModesInsideAndOnUnitCircle}.
Let us now show that $\Sigma_0-\Sigma_\infty\geq 0$.
If we use the system representation with correlated noise processes \eqref{eq:ClosedLoopDynamicsWithCorrelatedNoise}, the ARE for $\Sigma_\infty$, according to \citet{OptimalFiltering}, is
\begin{multline}
\label{eq:AlternativeAREForCorrelatedNoiseProcesses}
\Sigma_{\infty}=A_z^\prime \Sigma_\infty (A_z^\prime)^T +Q^\prime \\ - (A_z^\prime \Sigma_\infty C_z^T +S)(C_z\Sigma_\infty C_z^T + R)^{-1}(A_z^\prime \Sigma_\infty C_z^T +S)^T.
\end{multline}
Subtracting \eqref{eq:AlternativeAREForCorrelatedNoiseProcesses} from the Lyapunov equation for $\Sigma_0$ in Assumption~\ref{assum:SteadyStateClosedLoop} leads to
\begin{multline*}
\Sigma_0-\Sigma_\infty=A_z^\prime \big(\Sigma_0-\Sigma_\infty\big) (A_z^\prime)^T \\ + (A_z^\prime \Sigma_\infty C_z^T +S)(C_z\Sigma_\infty C_z^T + R)^{-1}(A_z^\prime \Sigma_\infty C_z^T +S)^T.
\end{multline*} 
This is also a Lyapunov equation with a unique solution since $\rho(A_z^\prime)<1$ (Assumption~\ref{assum:StableClosedLoop}). Further, we observe that
\begin{align*}
(A_z^\prime \Sigma_\infty C_z^T +S)(C_z\Sigma_\infty C_z^T + R)^{-1}(A_z^\prime \Sigma_\infty C_z^T +S)^T\geq 0,
\end{align*}
because $\Sigma_\infty\geq 0$. Therefore, we know that $\Sigma_0-\Sigma_\infty\geq 0$.
Hence, with initial condition $\Sigma_{0}$
\begin{align*}
\lim_{k\rightarrow\infty}\Sigma_z(k)=\Sigma_{\infty}
\end{align*}
if and only if $\rho(A_c)\leq 1$.
\end{pf}
\begin{corollary}
\label{cor:Problem1SolvedOptimal}
Problem~\ref{prob:ControllerStateEstimation} is solvable if and only if ${\rho(A_c)\leq 1}$.
\end{corollary}
Note that since the attacker uses a Kalman filter, it does not only obtain a perfect estimate of $x_c(k)$ but also an optimal estimate of $x(k)$. 

Theorem~\ref{thm:ConvergenceToSrongSolution} shows that the covariance matrix converges to the attacker's desired strong solution, but not how fast the convergence is. 
Therefore, we will now investigate the conditions for an exponential convergence rate.

\begin{proposition}
\label{prop:ExponentialConvergence}
Subject to $\Sigma_{0}>0$, the covariance matrix $\Sigma_z(k)$ converges exponentially fast to $\Sigma_{\infty}$ if and only if $\rho(A_c)<1$.
\end{proposition}
\begin{pf}
Theorem 4.1 in \citet{ConvergenceOfRiccatiDifferenceEquationWithSingularMatrices} shows us that subject to $\Sigma_{0}>0$ the covariance matrix $\Sigma_z(k)$ converges exponentially fast to the stabilizing solution if and only if $(C_z,A_z)$ is detectable and $(A_z,G)$ has no uncontrollable modes on the unit circle. 
We already showed that $(C_z,A_z)$ is detectable, therefore we look at the controllable modes of $(A_z,G)$ now.
Recall that $GG^T=Q$ such that 
\begin{align*}
G=\begin{bmatrix}
	\Sigma_w^{\frac{1}{2}} & 0 \\ 0 & 0
	\end{bmatrix}.
\end{align*}
For $(A_z,G)$ to have no uncontrollable modes on the unit circle we need $A_c$ to have no eigenvalues on the unit circle, because we cannot control the eigenvalues of $A_c$ with $G$, and due to Assumption~\ref{assum:ControlObserverAssumptions} $(A,\Sigma_w^{\frac{1}{2}})$ has no uncontrollable modes on the unit circle. 
We showed in Lemma~\ref{lem:StrongSolutionIffControllerModesInsideAndOnUnitCircle} that $\Sigma_\infty$ is a strong solution to the ARE if and only if $\rho(A_c)\leq 1$.  
Hence, subject to $\Sigma_{0}>0$ the covariance matrix $\Sigma_z(k)$ converges exponentially fast to $\Sigma_\infty$ if and only if ${\rho(A_c)<1}$.
\end{pf}
This shows us that if $\Sigma_{0}>0$ and the operator uses a stable controller, i.e., $\rho(A_c)<1$, the covariance matrix of the attacker's time-varying Kalman filter will converge exponentially fast to $\Sigma_{\infty}$. Hence, the attacker is able to obtain a perfect estimate of $x_c(k)$ exponentially fast.

\subsection{Breaking confidentiality of $x_c(k)$ using non-optimal observers}
Previously, we have shown under which conditions the attacker is able to get a perfect estimate of the controller state $x_c(k)$ when a time-varying Kalman filter is used. 
The time-varying Kalman filter is the optimal filter for linear systems with Gaussian noise.
One may wonder whether or not the attacker is able to perfectly estimate $x_c(k)$, when the attacker uses a non-optimal observer. Here, we investigate a time-invariant observer of the form
\begin{align}
	\label{eq:KalmanFilterAttackerTimeInvariant}
	\hat{z}(k+1)=A_z\hat{z}(k)+SR^{-1}y(k)+L_z\big(y(k)-C_z\hat{z}(k)\big),
\end{align}
with $\hat{z}(0)=0$, where $L_z$ is the attacker's constant observer gain.
As before, instead of looking at $\hat{z}(k)$, we analyse the error dynamics given by
\begin{align*}
	e_z(k+1)=\big(A_z-L_zC_z\big)e_z(k)+\eta(k)+L_zv(k).
\end{align*}
with $\mathbb{E}\lbrace e_z(k)\rbrace=0$ for all $k\geq 0$, covariance matrix $\mathbb{E}\big\lbrace e_z(k)e_z(k)^T \big\rbrace=\Sigma_z(k)$ and $\Sigma_z(0)\geq 0$.

The following theorem classifies all gains $L_z$ of a non-optimal observer such that Problem~\ref{prob:ControllerStateEstimation} is solved.
\begin{thm}
\label{thm:LinearTimeInvariantObserverGain}
For any $\Sigma_{z}(0)\geq 0$, 
\begin{align*}
\lim_{k\rightarrow\infty}\Sigma_z(k)=\tilde{\Sigma}_\infty=\begin{bmatrix}
\tilde{P} &0\\0 & 0
\end{bmatrix},
\end{align*}
if and only if $\rho(A_c)<1$, $L_z=[L_1^T\ 0^T]^T$ and $L_1\in \mathbb{R}^{n_x \times n_y}$ is chosen such that $\rho(A-L_1C)<1$. Here, $\tilde{P}$ is the unique solution to
\begin{align*}
\tilde{P}=(A-L_1C)\tilde{P}(A-L_1C)^T+\Sigma_w+L_1\Sigma_v L_1^T,
\end{align*}
and $\tilde{P}-P\geq 0$, where $P$ is the unique solution to \eqref{eq:RiccatiOperatorOneStep}.
\end{thm}
\begin{pf}
With $L_z=[L_1^T\ L_2^T]^T$ the error dynamics are
\begin{align*}
e_z(k+1)=\begin{bmatrix}
A-L_1C &BC_c\\
-L_2C & A_c
\end{bmatrix}e_z(k)+\begin{bmatrix}
w(k)-L_{1}v(k)\\ L_2v(k)
\end{bmatrix}.
\end{align*}
The error covariance matrix evolves as
\begin{multline}
\label{eq:ErrorCovarianceTimeInvariantCase}
\Sigma_z(k+1)=(A_z-L_zC_z)\Sigma_z(k)(A_z-L_zC_z)^T\\
+\begin{bmatrix}
\Sigma_w+L_1\Sigma_vL_1^T & L_1\Sigma_vL_2^T\\ L_2\Sigma_vL_1^T & L_2\Sigma_v L_2^T
\end{bmatrix}.
\end{multline}
Now we show that $\tilde{\Sigma}_\infty$ is the steady state solution of \eqref{eq:ErrorCovarianceTimeInvariantCase} if and only if $L_2=0$.
First, we observe that if $L_2=0$ then $\tilde{\Sigma}_\infty$ is a steady state solution of \eqref{eq:ErrorCovarianceTimeInvariantCase}, where $\tilde{P}$ is the solution to the Lyapunov equation
\begin{align*}
\tilde{P}=(A-L_1C)\tilde{P}(A-L_1C)^T+\Sigma_w+L_1\Sigma_v L_1^T.
\end{align*}
Note that $\tilde{P}\geq 0$ exists and is unique if ${\rho(A-L_1C)<1}$.
Second, if $\tilde{\Sigma}_\infty$ is a steady state solution of \eqref{eq:ErrorCovarianceTimeInvariantCase} the equations
\begin{align*}
\tilde{P}&=(A-L_1C)\tilde{P}(A-L_1C)^T+\Sigma_w+L_1\Sigma_vL_1^T, \\
0&=L_2(\Sigma_vL_1^T-C\tilde{P}(A-L_1C)^T),\ \mathrm{and} \\
0&=L_2(C\tilde{P}C^T+\Sigma_v)L_2^T
\end{align*}
are fulfilled.
The last equation is only fulfilled if $L_2=0$, since $\Sigma_v$ is positive definite. This simultaneously fulfils the second equation. The first equation recovers the Lyapunov equation for $\tilde{P}$.
Therefore, if $\tilde{\Sigma}_\infty$ is a steady state solution of \eqref{eq:ErrorCovarianceTimeInvariantCase} then $L_2=0$.
Hence, \eqref{eq:ErrorCovarianceTimeInvariantCase} has $\tilde{\Sigma}_\infty$ as a steady state solution if and only if $L_2=0$.
Let us now look at the convergence of \eqref{eq:ErrorCovarianceTimeInvariantCase} to $\tilde{\Sigma}_\infty$.
For any ${\Sigma_z(0)\geq 0}$, the error covariance matrix converges to $\tilde{\Sigma}_\infty$  if and only if $\rho(A_z-L_zC_z)<1$.
With $L_2=0$, the stability of $A_z-L_zC_z$ is guaranteed when both $\rho(A_c)<1$ and $\rho(A-L_1C)<1$. 
Due to detectability of $(C,A)$ in Assumption~\ref{assum:ControlObserverAssumptions} such a stabilizing $L_1$ exists.
Therefore, \eqref{eq:ErrorCovarianceTimeInvariantCase} converges to $\tilde{\Sigma}_\infty$ for any $\Sigma_z(0)\geq 0$, if and only if $L_2=0$, $\rho(A-L_1C)<1$, and $\rho(A_c)<1$.
Further, $\rho(A_z-L_zC_z)<1$ also makes $\tilde{\Sigma}_\infty$ the unique steady state solution of \eqref{eq:ErrorCovarianceTimeInvariantCase}. 
Since the Kalman filter is the best linear estimator, we know that $\tilde{P}-P\geq 0$ and $\tilde{P}=P$ if ${L_1=APC^T(CPC^T+\Sigma_v)^{-1}}$ \citep{OptimalFiltering}. This choice of $L_1$ turns the Lyapunov equation of $\tilde{P}$ into \eqref{eq:RiccatiOperatorOneStep}.
\end{pf}
Theorem~\ref{thm:LinearTimeInvariantObserverGain} shows us that the attacker is able to use the non-optimal observer \eqref{eq:KalmanFilterAttackerTimeInvariant} to solve Problem~\ref{prob:ControllerStateEstimation}, if and only if the controller is stable.
\begin{corollary}
\label{cor:Problem1SolvedNonOptimal}
Problem~\ref{prob:ControllerStateEstimation} is solvable with a non-optimal observer of the form \eqref{eq:KalmanFilterAttackerTimeInvariant} if and only if $\rho(A_c)< 1$.
\end{corollary}
According to Theorem~\ref{thm:LinearTimeInvariantObserverGain}, the attacker does not need to know the noise statistics $\Sigma_w$ and $\Sigma_v$ for the design of $L_1$ to estimate $x_c(k)$ perfectly, as long as $L_1$ is stabilizing. 
Hence, the attacker's required knowledge to solve Problem~\ref{prob:ControllerStateEstimation} is reduced when the operator uses a stable controller.
Further, the attacker has a smaller computational burden when a time-invariant observer is used.

\section{Defence mechanisms}
\label{sec:DefenceMechanism}
We presented under which conditions Problem~\ref{prob:ControllerStateEstimation} is solvable both with optimal and non-optimal strategies.
Therefore, we investigate now how to prevent the attacker from estimating $x_c(k)$ perfectly, i.e., make Problem 1 unsolvable.
We present a defence mechanism and discuss why an unstable controller is only in certain cases a good defence mechanism. 

\subsection{Injecting noise on the controller side}
As previously shown, an attacker under Assumption~\ref{assum:AttackerKnowledge} will be able to predict the controller state perfectly for $\rho(A_c)\leq1$.
We observe that the controller dynamics in \eqref{eq:ControllerDynamics} contain no uncertainty for the attacker when $y(k)$ is known.
Therefore, an approach for defence is to introduce uncertainty in the form of an additional noise term on the controller side.

The additional noise term $\nu(k)$ has a zero mean Gaussian distribution with a positive semi-definite covariance matrix $\Sigma_\nu\in\mathbb{R}^{n_c \times n_c}$.
Further, $\nu(k)$ is independent and identically distributed over time and also independent of $w(k)$, $v(k)$, and $z(0)$. 
The controller state with the additional noise term follows the dynamics
\begin{align*}
x_c(k+1)&=A_cx_c(k)+B_c y(k)+\nu(k).
\end{align*}
Here, $\nu(k)$ can be interpreted as process noise of the controller.

	\begin{assumption}
	The attacker knows the covariance matrix $\Sigma_{\nu}$ of the additional noise in the controller.
	\end{assumption}
	This assumption is in the spirit of Assumption~\ref{assum:AttackerKnowledge}, since the attacker has full model knowledge and knows the noise statistics of both $w(k)$ and $v(k)$.

This changes the process noise of the closed-loop system~\eqref{eq:closedLoopDynamics} from $\eta(k)$ to $\tilde{\eta}(k)=[w(k)^T\ \nu(k)^T]^T$
such that
\begin{align*}
\!\!\!\!\! \mathbb{E}\bigg\lbrace\begin{bmatrix}
\tilde{\eta}(k) \\ v(k)
\end{bmatrix}\begin{bmatrix}
\tilde{\eta}(k)^T & v(k)^T
\end{bmatrix}\bigg\rbrace
&=\left[
\begin{array}{cc|c}
\Sigma_w & 0 & 0\\ 0 & \Sigma_\nu & 0\\ \hline 0 & 0 & \Sigma_v
\end{array}
\right]
=\left[
\begin{array}{c|c}
\tilde{Q} &0\\\hline 0 &R
\end{array}
\right].
\end{align*}
The following proposition shows that with $\nu(k)$, the attacker's desired covariance matrix $\Sigma_\infty$ is not a steady state solution of \eqref{eq:RiccatiOneStepAheadARE} any more.
\begin{proposition}
\label{prop:InjectingNoiseMakesEstimationImpossible}
The algebraic Riccati equation~\eqref{eq:RiccatiOneStepAheadARE} with $Q=\tilde{Q}$ does \emph{not} have $\Sigma_{\infty}$ as a steady state solution.
\end{proposition}

\begin{pf}
With $\Sigma_z(k)=\Sigma_{\infty}$ and $Q=\tilde{Q}$ we obtain
\begin{align*}
	A_z\Sigma_{\infty}A_z^T+\tilde{Q}=
	\begin{bmatrix}
	APA^T+\Sigma_w &0\\
	0 &\Sigma_\nu
	\end{bmatrix},
\end{align*}
and using this in the Riccati equation~\eqref{eq:RiccatiOneStepAheadARE} leads to
\begin{align*}
\!\!\!\!\!\!\!	\Sigma_{\infty}=
	\begin{bmatrix}
	APA^T+\Sigma_w-APC^T(CPC^T+\Sigma_v)^{-1}CPA^T &0\\ 0 & \Sigma_\nu
	\end{bmatrix}.
\end{align*}
For $\Sigma_{\infty}$ to be a solution of \eqref{eq:RiccatiOneStepAheadARE} we need both
\begin{align*}
P=APA^T+\Sigma_w-APC^T(CPC^T+\Sigma_v)^{-1}CPA^T,
\end{align*}
which, as shown previously, exists, and $\Sigma_\nu=0$.

Since we assume $\Sigma_\nu\neq 0$, $\Sigma_{\infty}$ is not a solution of \eqref{eq:RiccatiOneStepAheadARE} any more.
\end{pf}
Here, we see that the attacker will not be able to perfectly estimate the controller's state if we use this additional noise on the controller side even if the attacker knows the noise properties.

Injecting $\nu(k)$ does not only lead to $\lim_{k\rightarrow\infty}\Sigma_z(k)=\tilde{\Sigma}_\infty\neq\Sigma_\infty$ as shown in Proposition~\ref{prop:InjectingNoiseMakesEstimationImpossible} but also changes the steady state covariance matrix of the closed-loop system from $\Sigma_0$ (see Assumption~\ref{assum:SteadyStateClosedLoop}) to $\tilde{\Sigma}_0$.
	The change in the covariance matrix, $\Delta\Sigma_0=\tilde{\Sigma}_0-\Sigma_0$, is given by
	\begin{align}
	\label{eq:LyapunovDeltaSigma0}
		\Delta\Sigma_0=A_z^\prime\Delta\Sigma_0(A_z^\prime)^T+\underbrace{\begin{bmatrix}
		0 & 0\\ 0 & \Sigma_{\nu}
		\end{bmatrix}}_{=\Delta Q}.
	\end{align}	
	Furthermore, we will quantify the performance degradation in the closed-loop system \eqref{eq:ClosedLoopDynamicsWithCorrelatedNoise} as $\mathrm{tr}(\Delta\Sigma_0)$, which represents the increase in total variation for the closed-loop system state.
	
	Therefore, we formulate the following convex optimization problem to determine $\Sigma_\nu$ subject to an upper bound $\gamma_p>0$ on the performance degradation.
	\begin{proposition}
	\label{prop:DetermineSigmaNu}
		The noise injection covariance $\Sigma_\nu$ that maximizes the controller confidentiality while keeping the performance degradation below a threshold, $\gamma_p>0$, is an optimal solution to the convex program,
		\begin{equation}
		\label{eq:OptimizationProblemToDetermineSigmaNu}
		\begin{aligned}
		&\max_{\Sigma_\nu,\tilde{\Sigma}_\infty}\ \mathrm{tr}(\tilde{\Sigma}_\infty)\\
		&\mathrm{s.t.} 
		\begin{cases}
				\begin{bmatrix}
				A_z\tilde{\Sigma}_\infty A_z^T+Q+\Delta Q-\tilde{\Sigma}_\infty & A_z\tilde{\Sigma}_\infty C_z^T\\
				\big(A_z\tilde{\Sigma}_\infty C_z^T\big)^T & C_z\tilde{\Sigma}_\infty C_z^T+R
				\end{bmatrix}\geq 0,\\
				\Delta\Sigma_0=A_z^\prime\Delta\Sigma_0(A_z^\prime)^T+\Delta Q\\
				\mathrm{tr}(\Delta\Sigma_0)\leq \gamma_p,\\
				\Sigma_{\nu}\geq 0,\quad
				\tilde{\Sigma}_\infty\geq 0,
		\end{cases}
		\end{aligned}		
		\end{equation}
		where $\Delta Q$ is given in \eqref{eq:LyapunovDeltaSigma0}.
	\end{proposition}
	\begin{pf}
		First note that both the objective and the constraints are convex in $\Sigma_\nu$ and $\tilde{\Sigma}_\infty$, which makes the optimization problem a convex semi-definite program and that problem \eqref{eq:OptimizationProblemToDetermineSigmaNu} is feasible, since $\Sigma_\nu=0$ and $\tilde{\Sigma}_\infty=\Sigma_\infty$ fulfil the constraints. 
		
		Next, the objective together with the first constraint guarantee that $\tilde{\Sigma}_\infty$ is the solution to the algebraic Riccati equation \eqref{eq:RiccatiOneStepAheadARE}, since the solution to \eqref{eq:RiccatiOneStepAheadARE} is the maximal solution to the algebraic Riccati inequality \citep{MaximalSolutionForRiccatiInequality}.
		
		Last, the second and third constraint impose the limitation on the allowed performance degradation, while the last two constraints enforce that the covariance matrices are positive semi-definite.
	\end{pf}
	
	\begin{remark}
	If a certain noise level in the controller is desired, the constraint $\Sigma_\nu\geq 0$ can be replaced by ${\Sigma_\nu\geq \gamma_cI_{n_c}}$, where $\gamma_c>0$. However, this tighter constraint can return an optimal $\tilde{\Sigma}_\infty$ with a smaller trace than in the case with $\gamma_c=0$. Furthermore, this constraint can make the problem infeasible, since a large $\gamma_c$ can interfere with the constraint on the performance bound.
	\end{remark}
	
\begin{remark}
The approach of adding some additional noise to the system is quite similar to the watermarking approach used, for example, in \citet{MoWatermarking}. 
The difference is that here the noise is added to the controller input, while in watermarking the noise is typically added to the output of the controller.
Therefore, these results show that if we position the watermarking noise at a different position we get the additional benefit of the attacker not being able to estimate the state of the controller perfectly.
\end{remark}

\subsection{An unstable controller as defence}
As shown before, Problem~\ref{prob:ControllerStateEstimation} is not solvable if and only if $\rho(A_c)>1$. 
Hence, designing the controller $(A_c,B_c,C_c,D_c)$ such that ${\rho(A_z^\prime)<1}$ and $\rho(A_c)>1$ leads to a successful defence against the discussed disclosure attack.

This implies that there are plants which have an inherent protection against the sensor attack. For example, all plants that are \emph{not} strongly stabilizable, i.e., plants that cannot be stabilized with a stable controller \citep{doyle2013feedback}, have an inherent protection against the estimation of the controller's state by the attacker.
Further, there are also control strategies that give an inherent protection to the closed-loop system. Disturbance accommodation control \citep{DisturbanceAccommodationControl}, where the controller tries to estimate a persistent disturbance, is one example of these control strategies.

If a plant can be stabilized by using a stable controller, i.e., a strongly stabilizing plant, using an unstable controller instead comes with several issues. 
A fundamental limitation is that the integral of the log sensitivity function is zero for a stable open-loop system. If the open-loop system has unstable poles the integral is equal to a constant positive value that depends on the unstable poles of the open-loop system and their directions for a multivariable discrete-time system \citep{BodeSensitivityMultivariableDiscreteTimeSystems}.
As \citet{RespectTheUnstable} shows with real world examples, it can have dire consequences if this fundamental limitation is not taken into account properly.
Hence, due to these fundamental limitation the introduction of unstable poles in the controller is not desirable.
Another issue of unstable controllers is that an unstable controller leads to an unstable open-loop system, if the feedback loop is interrupted.

Therefore, using an unstable controller for a strongly stabilizing plant is not recommended, but is an appropriate defence mechanism if an unstable controller is needed to stabilize the plant.

\section{Simulations}
In this section, we verify our results with simulations for a three-tank system. After stating the model of the three-tank system, we first show the effect of stable and unstable controllers on the attacker's estimate of the controller's state. 
Later, we verify that the additional noise prevents the attacker from estimating the controller's state perfectly.
\subsection{The three-tank system}
For the simulation of the closed-loop system estimation by the attacker we look at the following continuous-time three-tank system
\begin{align*}
	\dot{x}(t)&=
	\begin{bmatrix}
		-2 & 2 & 0\\
		2 & -4 & 2\\
		0 & 2 & -3
	\end{bmatrix}x(t)+
	\begin{bmatrix}
		0.5 & 0\\
		0 & 0\\
		0 & 0.5
	\end{bmatrix}u(t)+w(t),\\
	y(t)&=\begin{bmatrix}
	0 & 1 &0\\
	0 & 0 &1
	\end{bmatrix}x(t)+v(t).
\end{align*}
By discretizing the continuous-time system with a sampling period of ${T_s=0.5\,\mathrm{s}}$ we obtain $A$, $B$, and $C$.
We assume that $w(k)\sim\mathcal{N}(0,I_3)$ and $v(k)\sim\mathcal{N}(0,0.1I_2)$.

\subsection{Stable and unstable controllers}
Now that the system matrices are defined we are going to verify that the controller's stability influences the estimates of the controller's state by the attacker. We consider an observer-based feedback controller
\begin{align*}
x_c(k+1)&=(A-BK_i-LC)x_c(k)+Ly(k)\\
u(k)&=-K_ix_c(k)
\end{align*}
where $L$ is the observer gain and $K_i$ is the controller gain.
The closed-loop system matrix is then
\begin{align*}
	A_{z,i}^\prime=\begin{bmatrix}
	A &-BK_i\\ LC &A-BK_i-LC
\end{bmatrix}.
\end{align*}
According to Assumption~\ref{assum:StableClosedLoop}, ${\rho(A_{z,i}^\prime)<1}$, which means that $K_i$ and $L$ are designed such that $\rho(A-BK_i)<1$ and $\rho(A-LC)<1$.
The matrix $L$ is designed via pole placement to place the eigenvalues of $A-LC$ at $0.1$, $0.2$, and $0.3$.
Therefore, the error dynamics of the observer used in the controller are stable. In the following, we design three different $K_i$ such that $\rho(A-BK_i)<1$.

The first controller $K_S$ places the poles of $A-BK_S$ at $0.4$, $0.5$, and $0.6$. This first controller results in stable controller dynamics $A-BK_S-LC$ with $\rho(A-BK_S-LC)=0.4167$.

The second controller, $K_U$, is unstable, i.e., $\rho(A-BK_U-LC)>1$, but has no modes on the unit circle.
We determine $K_U$, such that $\rho(A-BK_U)< 1$ and $A-BK_U-LC$ has an eigenvalue at $1.5$.
The controller we obtain is
\begin{align*}
	K_U=\begin{bmatrix}
	0.5530 &   1.9589 &   1.2225\\
    1.8414 &  27.0785 & -12.9349
	\end{bmatrix}
\end{align*}
and it places the eigenvalues of $A-BK_U-LC$ at $1.5$, $-0.5175$, and $-0.1066$ and the eigenvalues of $A-BK_U$ at $0.6275$, $0.4272 + j0.6456$, and $0.4272 - j0.6456$.

For the design of the third controller, $K_I$, we place two eigenvalues inside the unit circle and one at 1, such that $\rho(A-BK_I-LC)=1$, while guaranteeing that $\rho(A-BK_I)<1$.
We obtain
\begin{align*}
K_I=\begin{bmatrix}
	 3.0988 &  -6.0472  &  2.3966 \\
     4.0471 &  10.8175  & -4.4516
\end{bmatrix},
\end{align*}
which places the eigenvalues of $A-BK_I-LC$ at $1$, $-0.2227$, and $-0.3693$ and the eigenvalues of $A-BK_I$ at $-0.2669$, $0.6405+j0.5942$, and $0.6405-j0.5942$.

For the first two controllers, the attacker designs a time-invariant Kalman filter with gain $L_{z}^{i}$ and steady state error covariance matrix $\Sigma_{\infty}^i=\lim_{k\rightarrow\infty}\Sigma^i(k)$, where $i\in \lbrace S,U\rbrace$.
The attacker's time-invariant Kalman filter design leads to an observer gain $L_z^S$ for the closed-loop system, which matches our results in Theorem~\ref{thm:LinearTimeInvariantObserverGain}. 
Since $K_U$ leads to an unstable controller, we know according to Corollary~\ref{cor:Problem1SolvedNonOptimal} that no time-invariant observer exists that solves Problem~\ref{prob:ControllerStateEstimation}. Further, Corollary~\ref{cor:Problem1SolvedOptimal} shows that even if the attacker would use a time-varying Kalman filter, Problem~\ref{prob:ControllerStateEstimation} is not solvable.

For the closed-loop system with $K_I$, the attacker needs to use a time-varying Kalman filter to obtain a perfect estimate of $x_c(k)$. 
The error covariance matrix in this case will converge to the same as in the case with $K_S$.

\begin{figure}
	\centering
	\includegraphics[scale=0.5]{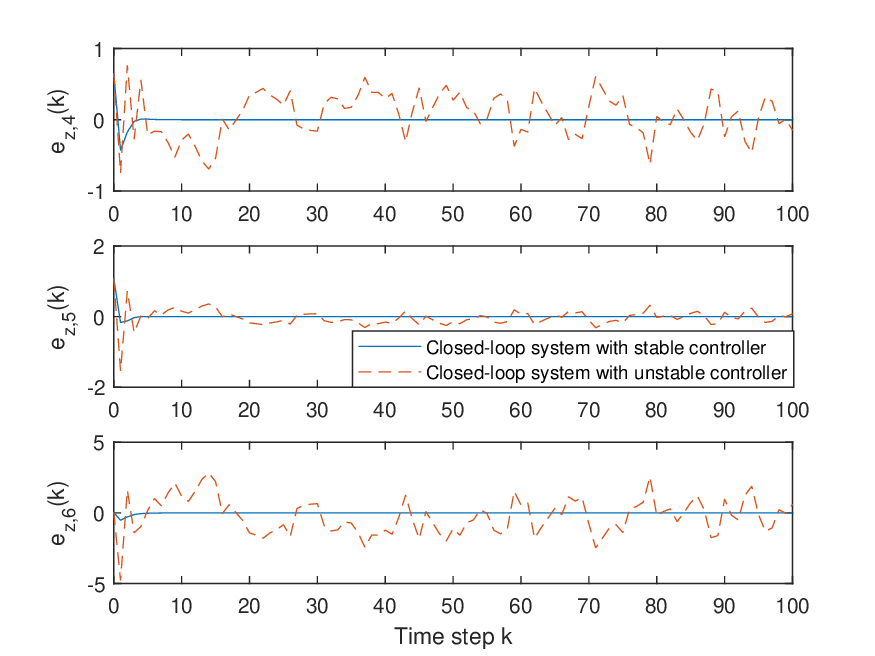}
	\caption{Comparison of the estimation error trajectories for the stable and unstable controller, $K_S$ and $K_U$ respectively}
	\label{fig:ErrorOfControllerState}
\end{figure}

Now that we designed the Kalman filters for each of the three closed-loop systems, let us look at the estimation error $e_z(k)=z(k)-\hat{z}(k)\in\mathbb{R}^6$. 
Here, we are only interested in the last three elements of $e_z(k)$, because they represent the estimation error of the controller state. The $j$th element of $e_z(k)$ is denoted by $e_{z,j}(k)$, where $j\in\lbrace 1, \cdots, 6 \rbrace$.
Figure~\ref{fig:ErrorOfControllerState} shows that in case of a stable controller $K_S$ the estimation error converges quickly to zero and the attacker obtains a perfect estimate of the controller's state.
However, if we use an unstable controller $K_U$ the estimation error remains noisy and the attacker is not able to obtain a perfect estimate of the controller's state.
\begin{figure}
\centering
\includegraphics[scale=0.5]{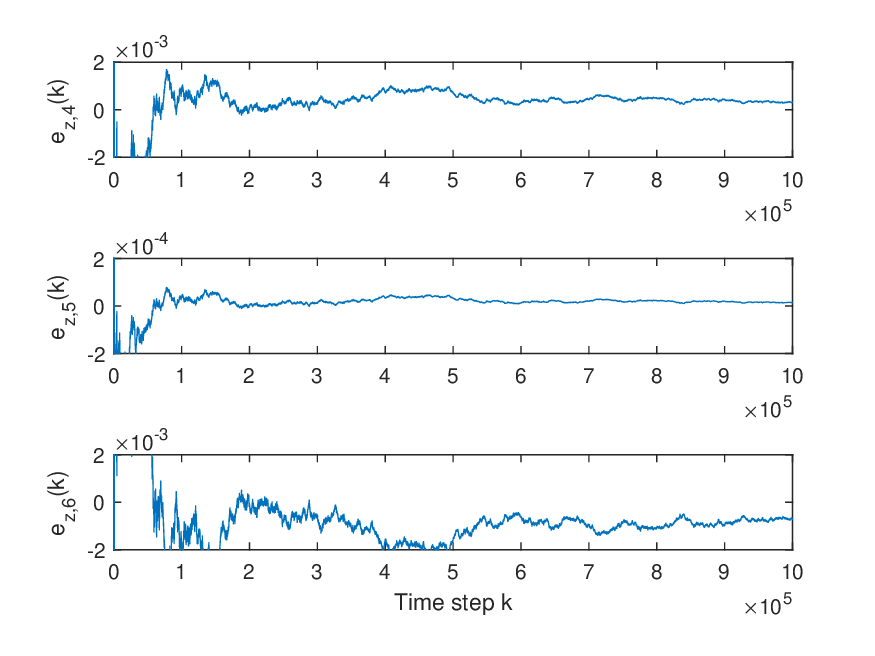}
\caption{Estimation error of the controller's state when the controller has a pole on the unit circle and the attacker uses a time-varying Kalman filter}
\label{fig:ErrorOfControllerStateWithEigenvalueOnUnitCircle}
\end{figure}
Furthermore, when $K_I$ is used, we observe that the estimation error converges to zero, but is still not zero after a million time steps (see Figure~\ref{fig:ErrorOfControllerStateWithEigenvalueOnUnitCircle}).
Theorem~\ref{thm:ConvergenceToSrongSolution} only tells us that the error will converge, but we know it does not converge exponentially by Proposition~\ref{prop:ExponentialConvergence}.
Although the attacker can obtain an almost perfect estimate with the time-varying Kalman filter after a million time steps, it is still not a perfect estimate. 
This shows us that a controller with modes on the unit circle can prevent the attacker from quickly obtaining a perfect estimate. 

\subsection{Injecting process noise for the controller}

Now that we showed how the controller design affects the attacker's estimate of the controller's state, we verify that injecting noise to the input of the controller prevents the attacker from estimating $x_c(k)$ perfectly. Further, we demonstrate how the choice of $\gamma_p$ affects both the attacker's estimate $\hat{x}_c(k)$, the plant's state $x(k)$, and the controller state $x_c(k)$.
	\begin{figure}
		\centering
		\includegraphics[scale=0.5]{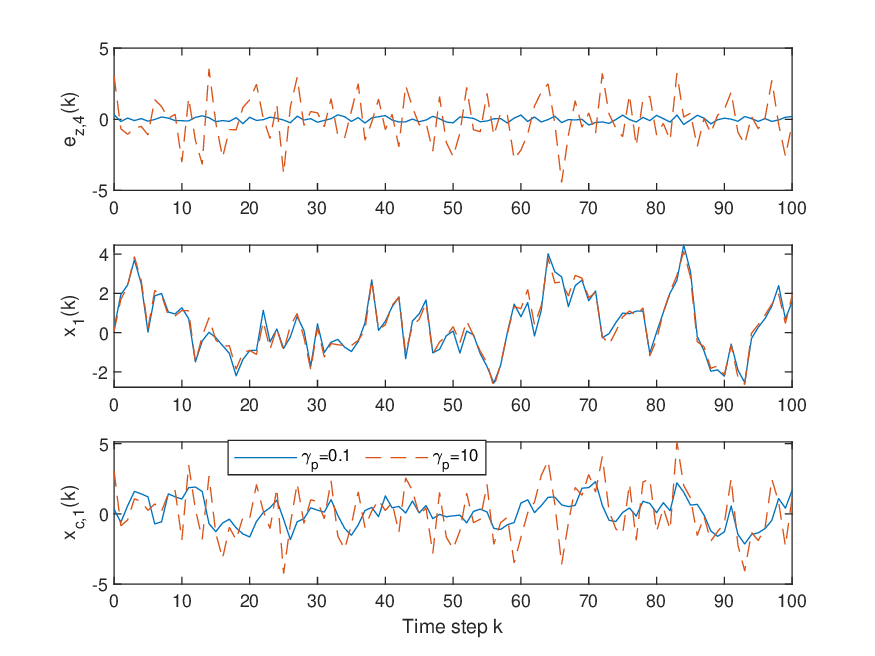}
		\caption{The effect of the additional noise on the first elements of the estimation error of the controller's state (upper plot), the plant's state (centre plot), and the controller's state (lower plot) when a stable controller is used}
		\label{fig:EstimationWithAdditionalNoise}
	\end{figure}	
	We investigate two cases for the performance degradation, one with a small allowed performance degradation, i.e., $\gamma_p=0.1$, and one with a large allowed performance degradation, i.e., $\gamma_p=10$.
	The operator uses the stable controller $K_S$ and the attacker uses again a time-invariant Kalman filter. 
	
	The upper plot of Figure~\ref{fig:EstimationWithAdditionalNoise} shows the trajectory of the attacker's estimation error of the first controller state. 
	Compared to Figure~\ref{fig:ErrorOfControllerState}, the estimation error exhibits noisy behaviour and the attacker is not able to obtain a perfect estimate even though the operator uses the stable controller $K_S$. Further, the noise around the estimation error increases the larger the allowed performance degradation is.
	
	To see the effect of the additional noise on the closed-loop system state, we show the trajectory of the first element of the plant's state, $x_1(k)$, in the centre plot and the trajectory of the first element of the controller's state, $x_{c,1}(k)$, in the lower plot of Figure~\ref{fig:EstimationWithAdditionalNoise}.
	We see that the state trajectory of the plant is not much more affected by the additional noise when we allow a hundredfold larger performance degradation, while the controller state becomes noisier. 
	
	The trajectories for the other elements of $x(k)$, $x_c(k)$, and the attacker's estimation error of the controller state behave similarly.
	Since the operator's objective is to control the plant optimally, this defence mechanism has the additional benefit of mostly increasing the noise in $x_c(k)$ but not considerably in the plant's state.
	
	Hence, the additional noise prevents the attacker from estimating the controller's state perfectly and additionally does not considerably affect the trajectory of the plant's state.

\section{Conclusion and future work}
We have shown exactly when an attacker with full model knowledge is able to perfectly estimate the internal state of an output-feedback controller by observing all measurements.

Although it seems obvious that an attacker according to our attack model can always estimate the controller's state, we gave necessary and sufficient conditions when an attacker is not able to obtain a perfect estimate. 
These conditions state that unstable controller dynamics prevent the attacker from obtaining a perfect estimate.
Further, the attacker can use a non-optimal time-invariant observer to perfectly estimate the controller state if and only if the controller has stable dynamics.
A defence mechanisms has been proposed to make the controller states confidential. 
This mechanism prevents the attacker from obtaining a perfect estimate by adding uncertainty to the controller dynamics. 
This is similar to watermarking approaches proposed by other authors with the twist that the noise signal is applied to the controller input and not to its output.
An unstable controller gives an inherent protection to plants that are not strongly stabilizable. However, designing such a controller introduces fundamental limitations on the sensitivity function of the closed-loop system and should only be used when an unstable controller is needed to stabilize the plant.

There are several directions of future work. 
It seems obvious that if the attacker has only access to a few sensors measurements, it will not be able to estimate the controller's state. 
However, it is interesting to investigate what happens if the attacker has access to some sensor and some actuator signals, and how many of each the attacker needs to get a perfect estimate of the controller's state.
Another research direction is to investigate the robustness of the controller state estimation for cases when the attacker has less model knowledge.

\bibliographystyle{plainnat}        
\bibliography{bibConfidentiality}           

\end{document}